\documentclass[runningheads]{llncs}
\usepackage{amsfonts}
\usepackage{epsfig}
\usepackage{amsmath}
\usepackage{graphicx}
\usepackage{amsmath}
\usepackage{amscd}
\usepackage{stmaryrd}
\usepackage{algorithm}
\usepackage{arydshln}
\usepackage{algorithmic}

\renewcommand{\algorithmicensure}{\textbf{Output:}} 
\usepackage{youngtab}
\usepackage{young}
\usepackage{geometry}
\usepackage{bbm}
\usepackage{multirow}
\usepackage{leftidx}

\usepackage[colorlinks,
            linkcolor=blue,
            anchorcolor=blue,
            citecolor=blue,
            ]{hyperref}

\def\bt{\begin{theorem}}
\def\et{\end{theorem}}
\def\bp{\begin{proposition}}
\def\ep{\end{proposition}}
\def\bc{\begin{corollary}}
\def\ec{\end{corollary}}
\def\bo{\begin{proof}}
\def\eo{\end{proof}}
\def\bx{\begin{example}}
\def\ex{\end{example}}
\def\br{\begin{remark}}
\def\er{\end{remark}}
\def\bl{\begin{lemma}}
\def\el{\end{lemma}}

\def\bn{\begin{definition}}
\def\en{\end{definition}}
\def\ba{\begin{array}}
\def\ea{\end{array}}
\def\be{\begin{equation}}
\def\ee{\end{equation}}
\def\bd{\begin{description}}
\def\ed{\end{description}}
\def\bu{\begin{enumerate}}
\def\eu{\end{enumerate}}
\def\bi{\begin{itemize}}
\def\ei{\end{itemize}}

\newbox\bigstrutbox
\setbox\bigstrutbox=\hbox{\vrule height12.5pt depth5pt width0pt}
\def\bigstrut{\relax\ifmmode\copy\bigstrutbox\else\unhcopy\bigstrutbox
\fi}

\newbox\Bigstrutbox
\setbox\Bigstrutbox=\hbox{\vrule height17.5pt depth5pt width0pt}
\def\Bigstrut{\relax\ifmmode\copy\Bigstrutbox\else\unhcopy\Bigstrutbox
\fi}

\def\ds{\displaystyle}

\def\A{{\bf A}}
\def\B{{\bf B}}
\def\C{{\bf C}}
\def\D{{\bf D}}
\def\E{{\bf E}}

\def\P{{\bf P}}

\def\x{{\bf x}}

\def\0{{\bf 0}}
\def\1{{\bf 1}}
\def\2{{\bf 2}}
\def\3{{\bf 3}}
\def\4{{\bf 4}}
\def\5{{\bf 5}}
\def\6{{\bf 6}}
\def\7{{\bf 7}}
\def\8{{\bf 8}}
\def\9{{\bf 9}}

\def\pnom{{pre\hbox{-}normal}}
\def\preR{{pre\hbox{-}\tilde{R}}}
\def\QJ{{{\mathbb Q}[\{x_l\,|\, l\in J\}]}}
\def\Rref{{Rre\hskip -.04cm f}}

\newcommand{\SWITCH}{\STATE \textbf{switch} }  
\newcommand{\ENDSWITCH}{\STATE \textbf{end switch}}  
\newcommand{\CASELINE}{\STATE \textbf{case} }  
\newcommand{\DEFAULTLINE}{\STATE \textbf{default } }

\begin{document}

\pagestyle{headings}

\mainmatter

\title{Riemann Tensor Polynomial Canonicalization by Graph Algebra Extension}

\titlerunning{Riemann Tensor Polynomial Canonicalization by Graph Algebra Extension}

\author{Hongbo Li,~~~~~, Zhang Li,~~~~~ Yang Li \\
hli@mmrc.iss.ac.cn,~~~~~lzshiw@mail.ustc.edu.cn,~~~~~1443188056@qq.com}

\authorrunning{Hongbo Li, Zhang Li, Yang Li}

\institute{Key Laboratory of Mathematics Mechanization, 
AMSS, UCAS, NCMIS, Chinese Academy of Sciences, 
Beijing 100190, China}

\maketitle

\begin{abstract}

Tensor expression simplification is an ``ancient" topic in computer algebra, a representative of
which is the canonicalization of Riemann tensor polynomials. 
Practically fast algorithms exist for monoterm canonicalization, but not for
multiterm canonicalization. Targeting the multiterm difficulty,
in this paper we establish the extension theory of graph algebra, and propose a 
canonicalization algorithm for Riemann tensor polynomials based on this theory.

\end{abstract}


\section{Introduction}


Tensor canonicalization
is a classical topic in computer algebra. There is a myriad of
softwares including this function, some of which are updating it consistently. An extensive collection of
the softwares and related literature can be found in  \cite{Martin-Garcia-xtensor}. 

The symmetries of the Riemann tensor on a Riemannian manifold of unknown dimension
are one of the most complex that occur in practice, making it a challenging task to normalize general
Riemann tensor polynomials. We adopt the following terminology in this paper:
\bd
\item[R-factor:]  indicial form of the Riemann tensor. It is of the form $R(ab,cd)$, where $a,b,c,d$ are {\it indices}.

\item[R-monomial:] scaled contraction of $n$ copies of the Riemann tensor; $n$ is called the {\it degree} of the R-monomial.


\item[Ricci R-factor:] 
an R-factor with {\it loop} (two dummy indices of the same name).

\item[R-polynomial:] a tensor whose indicial form is the sum of R-monomials.

\item[Row character:] the 4 indices of an R-factor can be written in two rows: the {\it upper} and {\it lower} ones.
\ed
The following are the symmetries within an R-monomial of degree $n$:
\bi
\item {\it $Sym8$-symmetry}: the group ${\mathbb Z}_2\times {\mathbb Z}_2\times {\mathbb Z}_2$ upon an R-factor;
it has two generators: $R(ab,cd)=-R(ba,cd)$ and $R(ab,cd)=R(cd,ab)$.

\item {\it Commutativity}: an R-monomial of degree $n$ has permutation symmetry $S_n$ among its R-factors.

\item {\it Renaming}: an R-monomial is invariant under renaming of dummy indices; it has permutation symmetry $S_d$, where
$d$ is the number of different dummy indices.

\item {\it Cyclic symmetry}: Bianchi identity $R(ab,cd)+R(ac,db)+R(ad,bc)=0.$
\ei
  
``$Sym8$" and ``Commutativity" form a group of size $8^n n!$, while ``Renaming" forms another group of size $d!$.
The two groups are called the {\it monoterm symmetry}, while the cyclic symmetry is called the
{\it multiterm symmetry}.

There is a symmetry on row character. Two indices of the same name can always invert their row characters.
This row symmetry is generally not taken into account, because a free index in any term of an R-polynomial
must have the same row character, a property determined by the tensor nature.

Suppose an order is given among the indices. Given an R-monomial $f$, 
in the equivalence class defined by the monoterm symmetry, an R-monomial whose sequence of indices has the minimal
lexicographic order is called the {\it pre-normal form} of $f$, also known as the canonical form of $f$ under
the monoterm symmetry. The canonical form of $f$ under all the symmetries of $f$ is called its {\it normal form} 
\cite{Fulling92}.

\vskip .2cm
We first introduce several methods in the literature devoted solely to computing the pre-normal form.

1. Renaming preference:

When all the permutations of dummy indices inside an R-monomial are listed, then sorting the
R-factors has logarithm complexity for any of the permutations. 
Any method of this class has factorial complexity \cite{Liu16}.

2. Double coset representative:

In \cite{Manssur02}, \cite{Manssur04}, a method based on strong generating set representation
of the permutation group is proposed, which has exponential complexity in the worst case \cite{Manssur02}.
The softwares \verb"xPerm" and \verb"Canon" based on
this algorithm turn out to be the fastest in practice \cite{Martin-Garcia08-2}.

3. Graph-theoretic method based on directed graph labeling:

A pair of dummy indices of a same name can be naturally taken as an edge connecting two vertices. 
To use the canonical form in graph theory, a formulation of 
tensor algebra as an {\it algebra of graphs} is proposed in \cite{Obeid01}.
In this thesis for Master's Degree, the first canonicalization
algorithm based on graph theory is proposed, where a tensor is formulated as a graph by representing
(1) its name as a vertex, (2) the indices as edges, (3) the row character of an index as direction of the edge, (4)
the position of an index by showing it on the edge. In canonical relabeling, the adjacency matrix 
is used to find the best permutation for identical vertices in a graph.
The algorithm should have factorial complexity.

4. Graph-theoretic method based on undirected graph labeling:

A different graph representation of tensors is proposed in \cite{Li16}, where 
(1) each index is a vertex, (2) each pair of identical dummy indices is an edge, (3) an extra vertex is constructed
to connect with every free index, (4) labels are used to describe the row character and position of
an index. A fast algorithm for graph isomorphism problem is used to 
find the canonical graph having the smallest sorted labeled edge set. 
The algorithm has factorial complexity in the worst case.

\vskip .2cm
The following methods are oriented to
computing the normal form instead of the pre-normal form only.

5. Group algebra and group representation theory:

In \cite{Ilyin91}, \cite{Ilyin96}, the group algebra of $Sym8\times S_n$ is used to reduce
computing the normal form to computing the orthogonal rejection from a linear subspace
in an surrounding space of dimension $8^n n!$. In \cite{Fulling92}, 
Young diagram and Schur program are used in normalization,
and are later implemented in \verb"Cadabra" \cite{Peeters}. 
These methods all have factorial complexity. In \cite{Kavian97}, a nondeterministic method
based on genetic algorithms is proposed.

6. Gr\"obner basis:

In \cite{Portugal98}, it is proposed that computing the normal form of an R-monomial can be
separated into two stages: first, computing the pre-normal form; second, 
using the Gr\"obner basis method to deal with multiterm symmetry.
To use the Gr\"obner basis method, all the permutations of dummy indices inside an
R-monomial must be considered. As the tensor contraction is highly restricted when
compared with polynomial multiplication, a Gr\"obner basis theory need be established
properly for tensor contraction \cite{Liu16}. The methods have factorial complexity. 

7. Linear equation solving among pre-normal forms related to cyclic symmetry.

\cite{Martin-Garcia07} introduces a method of normalization by
solving large systems of linear equations derived by alternatively 
replacing every R-factor in the pre-normal form of an R-monomial
with its 3-index antisymmetrization in all possible inequivalent ways.
The method has exponential complexity.


8. Term rewriting by graph structure analysis:

In \cite{Liu13}, a term rewriting method is proposed to normalize
a special class of R-monomials of degree 3
by analyzing the structure of the graph associated with each R-monomial.

As summarized in 
\cite{Martin-Garcia-xtensor}, there are 
fast monoterm canonicalization algorithms, while efficient
multiterm canonicalization algorithms are still missing. 
In some softwares a database 
storing the hard-to-compute normal forms of a large number of R-monomials is set up
for fast visit.
For example, 
\verb"MathTensor" \cite{Parker94} has an ever-growing list of \verb"RiemannRule"'s;
\verb"Invar" stores one million random Riemann monomial scalars and their
normal forms of degree ranging from 2 up to 100
\cite{Martin-Garcia07}, \cite{Martin-Garcia08-2}, \cite{Nutma14}.

\vskip .2cm
Targeting the difficulty in manipulating multiterm symmetry, in this paper:

i. We establish the {\it extension theory of graph algebra}.

The graph algebra proposed in \cite{Obeid01} cannot handle multiterm symmetry.
We first propose a new graph algebra by representing (1) an index as a vertex, with
the position in the inclusive R-factor as an intrinsic property 
of the vertex, (2) a pair of identical indices as an edge. 
Then we extend Grassmann's extension theory to the graph algebra,
where in an extension graph a vertex is a high-dimensional vector space spanned by
several vertices of the original graph. This framework makes it natural to 
handle cyclic symmetry and more general multiterm symmetry, where
algebraic manipulations are used in stead of graph-theoretic algorithms.

ii. We propose a easy-to-understand algorithm for pre-normal form computing, which has
the same worst-case complexity with the fastest algorithm 
\cite{Martin-Garcia08-2} in the literature.

The following observation is fundamental: when a configuration of R-factors is fixed, then to get the minimal
index sequence one only need fill in new dummy indices by the occurrence order of the positions held by old
dummy indices. It is called the ``positional order" in \cite{Balfagon98}, and has been implemented
in software \verb"TTC".


Our algorithm has no prerequisite of group theory.
Based on the above observation,
to fix a configuration our algorithm does not need run through the whole permutation group $S_n$. It has
worst-case complexity $O(n^2 2^n)$.

iii. 
We propose a complete algorithm for normal form computing based on graph algebra extension 
and linear equation solving in {\it rational functions field}.

There are two steps. 
First the pre-normal form of the extension graph is computed. Then
the normal form is computed by computing the rational numbers valued {\it RREF} (reduced row echelon form)
of a linear system with rational functions coefficient.
The complexity of the algorithm is $O(n9^n)$ by Gauss-Jordan elimination. In contrast, 
the ``brute-force" linear equation solving method \cite{Martin-Garcia07}
has complexity $O(27^n)$ if by Gauss-Jordan elimination. 

Throughout this paper we set the base numbers field to be the {\bf rational numbers $\mathbb Q$}, and set the order
among monomials/sequences to be the {\bf lexicographic order}.
This paper is organized as follows.
Sections 2 and 3 are on graph algebra extension theory;
Sections 4 and 5 are on algorithms for pre-normal form and normal form respectively;
Experiments and more examples will be reported elsewhere.

\section{Connection multigraph and detailed graph}

A {\it connection multigraph} is an undirected multigraph where the degree of any vertex $\leq 4$. Obviously
if the multigraph has more than one vertex, then any vertex has at most one loop. A vertex of degree $<4$ is
called a {\it free vertex}, while a vertex of degree 4 is
called a {\it dummy vertex}.
A dummy vertex with loop is called a {\it Ricci vertex}, while a dummy vertex without loop is called a
{\it complete vertex}. 

For any vertex $v$ of a connection graph $G$, associate with it
two twin-seats: $S_1S_2$ and $S_3S_4$. Now $v$ together with the two twin-seats
associated with it is called a {\it detailed vertex}, denoted by $v(S_1S_2, S_3S_4)$.

For any edge $e_j$ of $G$ connecting two vertices $v_1, v_2$, assign in each $v_i$ a seat
$s_{ij}$ so that $e_j$ connects the two seats $s_{1j}$ and $s_{2j}$, one from each detailed vertex.
If an assignment of all the edges to the seats changes $G$ into a graph
with the seats as vertices such that the degree of any seat $\leq 1$, then 
the new graph is called a {\it detailed graph} of $G$, and $G$ is called
the {\it detail-free multigraph} of the detailed graph.

A detailed graph $D$ can be formally multiplied with a scalar $\lambda\in \mathbb Q$, denoted by 
$\lambda D$, called a {\it multiple detailed graph}. 
Two multiple detailed graphs can be formally added, and if they are equal up to coefficient, they can be
combined by adding up their coefficients. The $\mathbb Q$-space spanned by finitely many 
detailed graphs is called the {\it detailed graph $\mathbb Q$-space} they generate.

Given a detailed graph $D$, the set of graphs with maximal vertex degree $\leq 1$ and containing $D$ as 
a subgraph span a $\mathbb Q$-space,
called the {\it ideal} generated by $D$.
For two detailed graphs $D_1, D_2$, 
a third detailed graph $D$ is said to be a {\it join} of
$D_1, D_2$, if both $D_1, D_2$ are subgraphs of $D$. Two detailed graphs can have more than one join.

We use ``$D_1D_2$" to denote a fixed join of $D_1, D_2$, and use ``$D_1D_2\cdots D_k$"
to denote a fixed join among $D_1, D_2, \ldots, D_k$. This formal product is commutative
and associative. For 
detailed graphs $D_1, D_2, \ldots, D_k$, 
we define the following join of $\lambda_1 D_1+\ldots+\lambda_{k-1} D_{k-1}$ and
$\lambda_k D_k$, called the {\it multilinear join}:
\be
(\lambda_1 D_1+\ldots+\lambda_{k-1} D_{k-1}) (\lambda_k D_k)
:=(\lambda_1\lambda_k) (D_1D_k)+\ldots+ (\lambda_{k-1}\lambda_k) (D_{k-1}D_k).
\ee
The multilinear join can be extended by associativity and commutativity to the general case.

Given finitely many detailed vertices, the formal scalar multiplication, addition and multilinear join
among them generate a finite dimensional $\mathbb Q$-space of detailed graphs, called the 
{\it detailed graph $\mathbb Q$-algebra} they generate. A general element of the algebra is called a 
{\it combined detailed graph}.

Given a detailed graph $D$ whose seats not connected by edges are said to be {\it free} and are labeled 
by {\it free indices} one by one,
if an order $O$ among its detailed vertices is given, then following the order all the seats 
of the detailed vertices can be lined up.
Along the sequence of seats $S$, if we remove all the free seats, and then preserve for each
edge only the first seat it connects, we get a sequence of seats $T$ where the serial number of each seat
is called the {\it dummy index} of the edge connecting it. Now label each seat of $S$ connected by an
edge with the dummy index of the edge. 
The resulting labeled sequence of $S$ is called the {\it serial index representation} of $D$
associated with the order $O$, also called the {\it positional order} in \cite{Balfagon98}.

In the serial index representation, the order among the free indices can be prescribed arbitrarily,
while the order among the dummy indices follow their natural order as integers. It is always assumed that
all free indices $\prec$ all dummy indices. For a detailed graph $D$ of $n$ detailed vertices,
there are $n!$ different orders among the detailed vertices, and consequently there are at most $n!$
different serial index representations of $D$. The minimal serial index representation in the lexicographic
order is called the {\it minimal index representation} of $D$, denoted by $\hbox{Min}_{I\hskip -.2mm dx}(D)$.

Given detailed graphs $D_1, \ldots, D_k$, their various serial index representations 
span a $\mathbb Q$-linear space, called the {\it serial index $\mathbb Q$-space} they generate.
For combined detailed graph $\sum_{i=1}^k \lambda_i D_i$, its {\it minimal index representation} refers
to $\sum_{i=1}^k \lambda_i \, \hbox{Min}_{I\hskip -.2mm dx}(D_i)$.

Given a detailed vertex $v(S_1S_2, S_3S_4)$, the following 24 detailed vertices
\be
\{v(S_{\sigma(1)}S_{\sigma(2)}, S_{\sigma(3)}S_{\sigma(4)})\,|\, \sigma\in S_4\}
\ee
span a $\mathbb Q$-space, called the {\it detail-free extension} of the detailed vertex.
It can be represented by the Grassmann exterior product of 24 detailed vertices
taken as vectors, if the vectors are linearly independent. No matter how
linearly dependent the 24 detailed vertices are, their common detail-free extension can be represented
unanimously in parametric form as follows:
\be
Ext(v(S_1S_2, S_3S_4)):=
\sum_{\sigma\in S_4} \lambda_\sigma v(S_{\sigma(1)}S_{\sigma(2)}, S_{\sigma(3)}S_{\sigma(4)}),
\ee
where the $\lambda$'s are free parameters. This form of representation is called the {\it detailed
extension} of the detailed vertex.

Given a detailed graph $D$ with $n$ detailed vertices $\{v_i(S_{i1}S_{i2}, S_{i3}S_{i4}))\,|\, i=1..n\}$,
or equivalently in the notation of detailed graph $\mathbb Q$-algebra, 
\be
D=\prod_{i=1}^n v_i(S_{i1}S_{i2}, S_{i3}S_{i4}), 
\ee
the corresponding detail-free multigraph among the $n$ detail-free extensions of the detailed
vertices can be represented as the following ``detailed graph"
among the $n$ detailed extensions of the detailed vertices:
\be\ba{lrl}
Ext(D) &:=& \ds
\prod_{i=1}^n 
\Big(\sum_{\sigma_i\in S_4} (\lambda_{\sigma_i} 
v_i(S_{\sigma_i(1)}S_{\sigma_i(2)}, S_{\sigma_i(3)}S_{\sigma_i(4)}))\Big) \\

&=& \ds\sum_{(\sigma_1, \ldots, \sigma_n)\in (S_4)^n} \hskip -.12cm
\Big(\prod_{i=1}^n \lambda_{\sigma_i}\Big)
\Big(\prod_{i=1}^n v_i(S_{\sigma_i(1)}S_{\sigma_i(2)}, S_{\sigma_i(3)}S_{\sigma_i(4)})\Big),
\ea
\label{def:ext}
\ee
called the {\it detailed extension} of $D$. Two detailed graphs are said to be {\it multigraph-like}, if they
have the same connection multigraph. 

So when viewed from the $\mathbb Q$-space spanned by detailed vertices,
the detail-free multigraph of a detailed graph $D$ with $n$ detailed vertices each having $m_i$-dimensional
detailed extension for $i=1..n$, is a multigraph whose vertices are 
$m_i$-dimensional linear subspaces for $i=1..n$; when 
viewed from the $\mathbb Q$-space spanned by detailed graphs,
the detail-free multigraph is a 
$(\prod_{i=1}^n m_i)$-dimensional linear subspace spanned by all the 
detailed graphs whose detailed vertices each have the same set of seats with
the corresponding detailed vertex of $D$. 

In an R-monomial, if an R-factor has all its dummy indices removed,
and all its free indices viewed as a set, then when the remainder is taken as a vertex,
and every pair of identical dummy indices is taken as an edge, 
a multigraph is obtained, called the {\it multigraph} of the R-monomial.

An R-monomial is said to be {\it connected} if so is its multigraph. 
The multigraph of a connected R-monomial is a connection multigraph.
For an R-monomial $f$, 
a connected R-submonomial $h$ is said to be {\it maximal} if there is no
connected R-submonomial of $f$ containing $h$ properly.
The multigraph of an R-monomial is the disjoint union of the connection multigraphs of all
the maximal connected R-submonomials.

For example, the connection multigraph of $R^{a\phantom{b,c}d}_{\phantom{a}b,c} R_{a}^{\phantom{a}e,fc}$ is 
$R_{\{b,d\}} \asymp R_{\{e,f\}}$, where each arc denotes an edge. The connection multigraph of 
$R^{a\phantom{b,}c}_{\phantom{a}b,\phantom{c}a}$ is $\stackrel{\frown}{R}_{\{b,c\}}$, while the connection multigraph of 
$R^{a\phantom{b,a}b}_{\phantom{a}b,a}$ is $\stackrel{\stackrel{\scriptstyle \frown}{\hbox{\it R}}}{\scriptstyle \smile}$.

In a connected R-monomial, every R-factor is a 4-tuple of seats, each seat being occupied by an index.  
When every seat is taken as a vertex, and every pair of identical dummy indices is
taken as an edge, a detailed graph is obtained with
every R-factor as a detailed vertex, called the {\it detailed graph} of the R-monomial. 
The {\it detailed graph} of an R-monomial is the disjoint union of the detailed graphs of
all the maximal connected R-submonomials.

For example, the detailed graph of 
$R^{\phantom{ba,}ac}_{ba,}$ is $R_{b*,} 
\hskip -.3cm\stackrel{/}{\phantom{\scriptscriptstyle 1}}^{*c}$, 
while the detailed graph of 
$R^{ab,}_{\phantom{ab,}ab}$ is $R^{\stackrel{**,}{\phantom{1}}}_{ \stackrel{ \phantom{1} }
{ \phantom{**}** }}
\hskip -.5cm \backslash \backslash$. Each edge is denoted by a line with two asterisk 
ends denoting the seats it connects. The serial index representations of the two detailed graphs
are $b11c, 1212$ respectively, which are also the minimal index representations of them.


\section{Detailed Pre-R-graph and Detailed R-graph}

Let $G$ be a connection multigraph. Denote by $Detail(G)$ the $\mathbb Q$-space of detailed graphs having the same
detail-free expansion $G$.

Define the following equivalence relation in $Detail(G)$:
two multiple detailed graphs 
$\mu_1 D_1$ and $\mu_2 D_2$ are equivalent if one of the following is satisfied:
\bd
\item[Sym$-$:] 
$\mu_1+\mu_2=0$, and $D_1$ and $D_2$ have only one different detailed vertex: 
if the detailed vertex of $D_1$ takes
the form $v(S_1S_2, S_3S_4)$, then the other in $D_2$ is $v(S_2S_1, S_3S_4)$.

\item[Sym+:] $\mu_1-\mu_2=0$, and $D_1$ and $D_2$ have only one different detailed vertex: if the detailed vertex in $D_1$ takes
the form $v(S_1S_2, S_3S_4)$, then the other in $D_2$ is $v(S_3S_4, S_1S_2)$.
\ed
Two combined detailed graphs $\sum_{i=1}^m \lambda_i D_{1i}$ and $\sum_{j=1}^m \mu_j D_{2j}$ are equivalent
if for $i=1..m$, $\lambda_i D_{1i}$ and $\mu_i D_{2i}$ are equivalent. This equivalence relation
is called the {\it pre-R-equivalence}. 
The pre-R-equivalence class 
of a detailed vertex $v$, detailed graph $D$, respectively, is
called a {\it detailed pre-R-vertex}, {\it detailed pre-R-graph}, and denoted by
$\preR(v)$, $\preR(D)$, respectively. 


{\bf Sym}$\pm$ generate the symmetry group $Sym8$.
From the viewpoint of graph algebra extension, depending on whether a detailed vertex $v$ has two loops or not, the 
corresponding detailed pre-R-vertex is a 4-D space or 8-D space spanned by the trajectory
of group $Sym8$ upon $v$,
called the {\it $Sym8$-subextension} of the detailed vertex.
A detailed pre-R-graph is thus a detailed graph whose detailed vertices are $Sym8$-subextensions of the
corresponding detailed vertices of any detailed graph in the pre-R-equivalence class.

Now that the coset $S_4/Sym8$ has 3 elements, the detailed extension of a loop-free detailed pre-R-vertex 
$v(S_1S_2, S_3S_4)$ is a 3-space spanned by
the following basis:
\be
v(S_1S_2, S_3S_4), \ \ \
v(S_1S_3, S_4S_2),\ \ \
v(S_1S_4, S_2S_3).
\label{pre:3}
\ee
The detailed extension of a detailed pre-R-vertex with loop is a 1-space spanned by itself.

For a detailed pre-R-graph 
\be\ba{l}
D=\prod_{i=1}^n v_i(S_{i1}S_{i2}, S_{i3}S_{i4}), \ \hbox{ where }\\
\hbox{the }v_i \hbox{ for } i\leq r \hbox{ have loop, and for }i>r \hbox{ are loop-free},\bigstrut
\label{ex:loop:no}
\ea
\ee
the detailed extension $Ext(D)$ is a $3^{n-r}$-dimensional $\mathbb Q$-space with the following parametric representation:
\be\ba{lll}
Ext(D) &=& \ds
\Big(\lambda_r \prod_{i=1}^r v_i(S_{i1}S_{i2}, S_{i3}S_{i4})\Big)
\Big(\prod_{i=r+1}^n 
(\lambda_{i2}v_i(S_{i1}S_{i2}, S_{i3}S_{i4})
+\lambda_{i3}v_i(S_{i1}S_{i3}, S_{i4}S_{i2})
+\lambda_{i4}v_i(S_{i1}S_{i4}, S_{i2}S_{i3}))
\Big)\\

&=& \ds \sum_{i=r+1}^n\ \sum_{\sigma_i\in BP(1,2)}
\Big(\lambda_r \prod_{i=r+1}^n {\rm sign}(\sigma_i) \lambda_{i\sigma_i(2)}\Big) \\

&& \ds
\Big(\prod_{i=1}^r v_i(S_{i1}S_{i2}, S_{i3}S_{i4})\Big)
\Big(\prod_{i=r+1}^n v_i(S_{i1}S_{i\sigma_i(2)}, S_{i\sigma_i(3)}S_{i\sigma_i(4)})\Big),
\label{pre:set}
\ea
\ee
where $BP(1,2)$ is the set of bipartitions of 2,3,4 into two subsequences of length 1,2 respectively.
The basis in (\ref{pre:set}) is the set of all detailed pre-R-graphs having the same connection multigraph with $D$.


For a detailed graph $D\in Detail(G)$, 
the minimum in the lexicographic order of all the minimal serial 
representations of elements in $\preR(D)$ is called the {\it pre-normal form} of $D$, denoted by $\pnom(D)$.
For combined detailed graph $\sum_{i=1}^m \lambda_i D_i$ in the detailed graph $\mathbb Q$-space, 
its {\it pre-normal form} is $\sum_{i=1}^m \lambda_i\ \pnom(D_i)$.
The pre-normal form provides a unique 1-D representation of the detailed pre-R-graph.

Define the following equivalence relation in $Detail(G)$:
let $D_1, D_2, D_3$ be detailed graphs, and let
$\mu_1, \mu_2, \mu_3\in \mathbb Q$, then
$\mu_1 D_1$ and $\mu_2 D_2+\mu_3 D_3$ are equivalent if one of the following is satisfied:
\bd
\item[Pre-R:] $\mu_2 D_2+\mu_3 D_3$ and $\mu_1 D_1$ are pre-R-equivalent.

\item[Bianchi:] $-\mu_1=\mu_2=\mu_3$, and $D_1, D_2, D_3$ have only one different detailed vertex: 
if the detailed vertex in $D_1$ takes
the form $v(S_1S_2, S_3S_4)$, then the other two in $D_2, D_3$ separately are
$v(S_1S_3, S_4S_2)$, $v(S_1S_4, S_2S_3)$ respectively.
\ed
Two combined detailed graphs $\sum_{i=1}^m \lambda_i D_{1i}$ and $\sum_{j=1}^{2m} \mu_{j} D_{2j}$ of $Detail(G)$ are equivalent
if for $i=1..m$, $\lambda_i D_{1i}$ and $\mu_{2i} D_{2i}+\mu_{m+i} D_{2(m+i)}$ are equivalent. This equivalence relation
is called the {\it R-equivalence}. 

The R-equivalence class of a detailed vertex $v$, detailed graph $D$, respectively, is
called a {\it detailed R-vertex}, {\it detailed R-graph}, respectively, and denoted by 
$\tilde{R}(v)$, $\tilde{R}(D)$, respectively.
The R-equivalence relation naturally induces an equivalence relation among the detailed pre-R-graphs,
also called the {\it R-equivalence}: two detailed pre-R-graphs are equivalent if as detailed graphs
they are R-equivalent.

{\bf Bianchi} defines only one linear relation among the basis elements of (\ref{pre:3}). The reason is as follows.
If the 4 seats of a vertex are permuted, then there are as many as 24 Bianchi relations:
\be
v(S_{\sigma(1)}S_{\sigma(2)}, S_{\sigma(3)}S_{\sigma(4)})
+v(S_{\sigma(1)}S_{\sigma(3)}, S_{\sigma(4)}S_{\sigma(2)})
+v(S_{\sigma(1)}S_{\sigma(4)}, S_{\sigma(2)}S_{\sigma(3)})=0,
\ee 
for all $\sigma\in S_4$. It is easy to see that all of them are pre-R-equivalent.

So the detailed extension of a loop-free detailed R-vertex 
$v(S_1S_2, S_3S_4)$ is a 2-space spanned by
the following basis:
$v(S_1S_2, S_3S_4)$,
$v(S_1S_3, S_4S_2)$.
For detailed R-graph (\ref{ex:loop:no}),
$Ext(D)$ is a $2^{n-r}$-dimensional $\mathbb Q$-space with the following basis:
\be
\Big\{
\Big(\prod_{i=1}^r v_i(S_{i1}S_{i2}, S_{i3}S_{i4})\Big)
\Big(\prod_{i=r+1}^n v_i(S_{i1}S_{i\sigma_i(2)}, S_{i\sigma_i(3)}S_{i4})\Big)
\,\Big|\, 
\sigma_i\in S_2 \hbox{ acting upon }2,3 \Big\}.
\label{R:set}
\ee
(\ref{R:set}) is the set of all detailed R-graphs having the same connection multigraph with $D$.

For $D\in Detail(G)$, 
the minimum in the lexicographic order of all the pre-normal forms
of elements in $\tilde{R}(D)$ is called the {\it R-normal form} of $D$, denoted by $R$-$normal(D)$.
For combined detailed graph $\sum_{i=1}^m \lambda_i D_i$, 
its {\it R-normal form} is $\sum_{i=1}^m \lambda_i\ R\hbox{-}normal(D_i)$.
The R-normal form provides a unique 1-D representation of the detailed R-graph $\tilde{R}(D)$, or equivalently
the common multigraph $G$. 

For detailed graph $D$ in (\ref{ex:loop:no}), as a detailed pre-R-graph it has a 
detailed extension of $3^{n-r}$ dimensions, while as a detailed R-graph its detailed extension has dimension
$2^{n-r}$. So the $3^{n-r}$ basis elements in (\ref{pre:set}) when taken as detailed R-graphs satisfy
$3^{n-r}-2^{n-r}$ linear constraints. These constraints can be selected as the following $3^{n-r}-2^{n-r}$ 
equations: for all $r<i\leq n$, all
$\sigma_j \in S_2$ acting upon 2,3, and all $\sigma_k \in BP(1,2)$ acting upon 2,3,4,
\be\ba{l}
 \ds\pnom \Big\{ 
\Big(\prod_{s\leq r} v_s(S_{s1}S_{s2}, S_{s3}S_{s4}) \Big)
\Big(\prod_{r<j<i}
v_j(S_{j1}S_{j\sigma_j(2)}, S_{j\sigma_j(3)}S_{j4})\Big)\Bigstrut\\

 \ds \phantom{\pnom \Big\{}
\Big(v_i(S_{i1}S_{i2}, S_{i3}S_{i4})
+v_i(S_{i1}S_{i3}, S_{i4}S_{i2})+v_i(S_{i1}S_{i4}, S_{i2}S_{i3})\Bigstrut
\Big)\\

 \ds \phantom{\pnom \Big\{}
\Big(\prod_{k>i}
v_k(S_{k1}S_{k\sigma_k(2)}, S_{k\sigma_k(3)}S_{k\sigma_k(4)})\Big)\Bigstrut \Big\}
\
= 0.
\ea
\label{triangulate}
\ee
(\ref{triangulate}) can also be obtained from the pre-normal form of the detailed 
extension (\ref{pre:set}) by $3^{n-r}-2^{n-r}$ special evaluations of the $\lambda$'s.

(\ref{triangulate}) is a linear homogeneous system of $3^{n-r}-2^{n-r}$ equations in 
$m\leq 3^{n-r}-2^{n-r}$ unknowns, where each unknown is a pre-normal R-monomial. 
Denote the unknowns by $x_1\succ x_2\succ  \cdots \succ  x_m$ 
following the lexicographic order. 
Let $\E$ be the {\bf RREF} (reduced row echelon form) of the coefficient matrix.
If $\pnom(f)$ is up to coefficient a leading variable of an equality in $\E(x_1, x_2, \ldots, x_m)^T=0$, then 
the normal form of $f$ is obtained by substituting the equality into $\pnom(f)$, else
$\pnom(f)$ is the normal form of $f$.

\section{Algorithm for Pre-normal Form}

For an R-monomial $f$, its {\it pre-normal form} is an R-monomial $g$ whose 
index sequence is the pre-normal form of the detailed graph of $f$. For an R-polynomial, its 
{\it pre-normal form} is the linear combination of the pre-normal forms of its terms.
An R-polynomial is said to be {\it pre-normal} if it is its own pre-normal form. 
For example,
the pre-normal form of an R-factor $R(ij,kl)$ can be obtained in three steps: (1) sort
$i,j$ non-decreasingly, (2) sort $k,l$ non-decreasingly, (3) sort the two sorted pairs non-decreasingly.

We assume that the input R-polynomial does not contain indices named after integers, so that we can introduce
integers as new dummy indices. We always assume 
\[
\hbox{free indices} \prec \hbox{new dummy indices (integers)} \prec \hbox{input dummy indices}.
\]
Before introducing the pre-normal form computing algorithm, let us check three typical examples.

\vskip .2cm
{\it Example 1}. Let $f=R_{d_1\phantom{d_2,d_6}d_7}^{\phantom{d_1}d_2,d_6} 
R_{\phantom{d_3d_4,d_7} d_6}^{d_3d_4,d_7} 
R^{d_1d_5,}_{\phantom{d_1d_5,}d_2a} 
R^{b}_{\phantom{b}d_4,d_3d_5}$, where
$a\prec b\prec d_1\prec \ldots \prec d_7$.
\vskip .2cm

First, change each R-factor into its pre-normal form. The result is
$R(d_1d_2,d_6d_7)$, $-R(d_3d_4,d_6d_7)$, $-R(ad_2,d_1d_5)$, $R(bd_4,d_3d_5)$.

Second, the R-factors are classified into two groups: the first group $Q_F=\{-R(ad_2,d_1d_5), R(bd_4,d_3d_5)\}$
consists of R-factors having free indices, the second group $Q_D=\{R(d_1d_2,d_6d_7), -R(d_3d_4,d_6d_7)\}$ consists of the rest.
Obviously all elements of $Q_F \prec$ all elements of $Q_D$.

Third, 
$Q_F$ is sorted by free indices: $-R(ad_2,d_1d_5)\prec R(bd_4,d_3d_5)$, 
making $Q_F$ a sequence. The serial index representation of $Q_F$ is then obtained by the following assignment:
\[
d_2 \rightarrow 1,\ \
d_1d_5 \rightarrow 23,\ \
d_4\rightarrow 4,\ \
d_3\rightarrow 5.
\]
The assignment naturally branches into two options: (1) $d_1 \rightarrow 2,\ d_5 \rightarrow 3$;
(2) $d_1 \rightarrow 3,\ d_5 \rightarrow 2$. Within sequence $Q_F$, option (1) gives index sequence $-a123,-b435$ while
option (2) gives index sequence $-a123,-b425$. As option (2) gives lower order, it becomes the single option.

Fourth, the corresponding old dummy indices in group $Q_D$ are
also renamed by the new ones. While $Q_F$ is updated to $-R(a1,23), -R(b4,25)$, 
$Q_D$ is updated to $\{-R(13,d_6d_7), R(45,d_6d_7)\}$. Obviously $-R(13,d_6d_7)\prec R(45,d_6d_7)$.
Now that all elements of $Q_F$ and $Q_D$ are ordered, they can merge to form a single sequence. In other words,
the order among the 4 input R-monomials have been fixed. 

Finally, the remaining old dummy indices
$d_6d_7$ are assigned to new ones $67$. No matter whether $d_6$ is renamed to 6 or 7, the resulting R-monomial
is the same: $-R(a1,23) R(b4,25) R(13,67) R(45,67)$. It is the pre-normal form of $f$.

\vskip .2cm
Example 1 shows that if $f$ has free indices, then its pre-normal form can be computed by
the loop procedure of first sorting the R-factors having fixed indices,
then renaming the old dummy indices in the sorted R-factor sequence
by the serial index representation. 
The same idea applies to the case when there is no free index.

\bl \label{lem:2}
If connected R-monomial $f$ has no free index but has at least one Ricci R-factor, 
then the pre-normal form of $f$ must be led by a Ricci R-factor. 
\el

{\it Proof}. Any Ricci R-factor in the first position can be renamed as $R(11,23)$ or $R(12,13)$,
while a non-Ricci R-factor has the minimal index form $R(12,34)$, which is higher in order. 
\hfill $\square$

\vskip .2cm
{\it Example 2}. $f=R_{d_1\phantom{d_2,d_3}d_4}^{\phantom{d_1}d_2,d_3} R^{d_5d_4,}_{\phantom{d_5d_4,}d_5d_3} 
R_{d_2d_6,}^{\phantom{d_2d_6,}d_1d_6}$.
\vskip .2cm

The R-factors in pre-normal form are $R(d_1d_2,d_3d_4), R(d_3d_5,d_4d_5), R(d_1d_6,d_2d_6)$. 
By Lemma \ref{lem:2}, one of the two Ricci R-factors is the first in the pre-normal form of $f$. For example,
if $R_{d_3d_5,d_4d_5}$ is the first, then $d_5\rightarrow 1$ and $d_3d_4\rightarrow 23$. So 
4 branches are generated in assigning new indices 1,2,3:
\[\ba{llll}
(1) & d_5\rightarrow 1, & d_3\rightarrow 2, & d_4\rightarrow 3; \\
(2) & d_5\rightarrow 1, & d_3\rightarrow 3, & d_4\rightarrow 2; \\
(3) & d_6\rightarrow 1, & d_1\rightarrow 2, & d_2\rightarrow 3; \\
(4) & d_6\rightarrow 1, & d_1\rightarrow 3, & d_2\rightarrow 2.
\ea\]

In each branch, the sequence of R-monomials containing fixed indices is denoted by $Q_F$, and the other R-monomials
are in the set $Q_D$:
\[\ba{llll}
(1) & Q_F=[ \phantom{-}R(12, 13), \phantom{-} R(23,d_1d_2)], & Q_D=\{R(d_1d_6,d_2d_6)\}; \\

(2) & Q_F=[-R(12, 13),  \phantom{-}R(23,d_1d_2)], & Q_D=\{R(d_1d_6,d_2d_6)\}; \\

(3) & Q_F=[ \phantom{-}R(12, 13),- R(23,34)], & Q_D=\{R(d_3d_5,d_4d_5)\}; \\

(4) & Q_F=[ \phantom{-}R(12, 13),  \phantom{-}R(23,34)], & Q_D=\{R(d_3d_5,d_4d_5)\}.
\ea
\]

In branch (1), assignment $d_1d_2\rightarrow 45$ generates one more branch: while
option $d_1\rightarrow 4, d_2\rightarrow 5$ leads to 
$R(12, 13)R(23,45)R(4d_6,5d_6)$, option $d_1\rightarrow 5, d_2\rightarrow 4$ leads to 
$-R(12, 13)R(23,45)R(4d_6,5d_6)$. Since the two differ by coefficient, $f=0$.

\vskip .2cm
Example 2 shows that depth-first strategy is preferred in generating branches.

\vskip .2cm
{\it Example 3}. Set $a=b$ in Example 1. 

\vskip .2cm
The R-factors of $f$ are 
$R(d_1d_2,d_6d_7)$, $-R(d_3d_4,d_6d_7)$, $-R(ad_2,d_1d_5)$, $R(ad_4,d_3d_5)$.
Any of them may be the first in the pre-normal form of $f$. For example,
if $R(d_1d_2,d_6d_7)$ is the first, then assignment $d_1d_2d_6d_7 \rightarrow 1234$ has 24 options.
All together the assignment of new indices 1,2,3,4 generates $4\times 24=96$
branches. 


Once the first new indices are assigned, then similar to Example 1, more new indices can be assigned
in each branch, and new branches may be generated by different options in the new assignment.
Each branch finally generates a serial index representation of $f$, and the minimum of them
gives the pre-normal form of $f$. By the algorithm below, the pre-normal form is 
$R(12,34) R(12,56) R(37,48) R(57,68)$.

An R-factor is said to be {\it free}
if it contains any free index, otherwise it is said to be {\it dummy}. 
A dummy R-factor is said to be {\it complete} if it is not Ricci.

In the following,
\bi
\item
$J$ records the number of branches generated, 
\item
$K$ records the serial number of the current branch in process, 
\item
$I[k]-1$ records the number of different new indices introduced in the $k$-th branch, 
\item
$Q_F[k]$ records the sequence of R-factors having fixed indices in the $k$-th branch,
\item
$Q_D[k]$ records the set of R-factors not in $Q_F[k]$,
\item
$V[k]$ records the set of fixed indices in the $k$-th branch.
\ei

\begin{algorithm}[H]
\caption{ ``\texttt{pnom}":\ \ \ computing pre-normal form of connected R-monomial}

\begin{algorithmic}[1]
\REQUIRE
a connected R-monomial $f$ whose indices are not integers; \\

an order where all free indices $\prec$ all integer indices $\prec$ all input dummy indices.

\ENSURE
$p$, the pre-normal form of $f$

\STATE 

Set $E:=$ sequence of free R-factors of $f$, sorted in increasing order.\\

Set $L:=$ set of dummy Ricci R-factors. \\

Set $C:=$ set of complete R-factors. 

\STATE

Set $f:=$ sequence of R-factors of $f$ each in pre-normal form.

If any element of $f$ is of the form $\lambda R(b_1b_1,b_2b_3)$ or $\lambda R(b_1b_2,b_3b_3)$, return 0 and exit. \\

If $L\cup C=\emptyset$, return $f$ and exit.

\STATE Set $p:=f$, set $K:=1$.

\SWITCH ($E,L$)

\CASELINE $E\neq \emptyset${\bf :}\ \ \ (case of Example 1)

\hskip .6cm Set $J=K=I[1]:=1$, set $V[1]:=$ set of free indices of $E$.

\hskip .6cm Set $Q_F[1]:=E$, set $Q_D[1]:=L\cup C$. \\

\hskip .6cm Execute procedure \texttt{SerIdx}.

\CASELINE $E=\emptyset${\bf ,} $L\neq \emptyset${\bf :} \ \ \ (case of Example 2)

\hskip .6cm {\bf switch} (${\rm deg} (f)$)\\

\hskip .6cm {\bf case $=1$:}
\ \ \ ($f$ must be in the form $\lambda R(b_1b_2,b_1b_2)$)

\hskip 1.2cm Return $\lambda R(12,12)$ and exit.

\hskip .6cm {\bf case $>1$:}

\hskip 1.2cm Set $J:=2\times \#L$.\\
 
\hskip 1.2cm {\bf for} $i$ from 1 to $\#L$, let $L[i]$ be the $i$-th element of $L$, {\bf do  }

\hskip 1.8cm Let $b_1$ be the index of multiplicity 2 in $L[i]$, and $b_2, b_3$ 
be the other two indices. 

\hskip 1.8cm  Set $Q_F[2i-1]:=$ pre-normal form of $L[i]$ after renaming 
$b_1, b_2, b_3$ as $1, 2, 3$ respectively.

\hskip 1.8cm  Set $Q_D[2i-1]:=L\cup C\backslash \{L[i]\}$ after the above renaming. 

\hskip 1.8cm  
Set $Q_F[2i]:=$ pre-normal form of $L[i]$ after renaming 
$b_1, b_3, b_2$ as $1, 2, 3$ respectively.

\hskip 1.8cm  Set $Q_D[2i]:=L\cup C\backslash \{L[i]\}$ after the above renaming. 

\hskip 1.8cm  Set $V[2i-1]=V[2i]:=\{1, 2, 3\}$, set $I[2i-1]=I[2i]:=4$.

\hskip 1.2cm {\bf end for}

\hskip 1.2cm Execute procedure \texttt{SerIdx}.

\hskip .6cm {\bf end switch}

\DEFAULTLINE \ \ \ (case of Example 3, $E=L=\emptyset$)













\hskip .6cm Set $J:=24\times {\rm deg}(f)$.

\hskip .6cm {\bf for} $i$ from 1 to ${\rm deg}(f)$, let the $i$-th element $C[i]$ of $C$ be of the form $\lambda R(b_1b_2,b_3b_4)$,
{\bf do}

\hskip 1.2cm {\bf for} all permutation
$\sigma_j\in S_4$, where $j=1..24$, {\bf do}

\hskip 1.8cm Set $h=24(i-1)+j$. 

\hskip 1.8cm Set $Q_F[h]:=$ pre-normal form of $C[i]$ after renaming 
$b_1, b_2, b_3, b_4$ as $\sigma_j(1), \sigma_j(2), \sigma_j(3), \sigma_j(4)$ respectively.

\hskip 1.8cm Set $Q_D[h]:=C\backslash \{C[i]\}$ after the above renaming. 

\hskip 1.8cm  
Set $V[h]:=\{1, 2, 3, 4\}$,  set $I[h]:=5$.

\hskip 1.2cm {\bf end for}

\hskip .6cm {\bf end for}

\hskip .6cm Execute procedure \texttt{SerIdx}.


\ENDSWITCH

\RETURN $p$.

\end{algorithmic}
\end{algorithm}

\renewcommand{\algorithmicensure}{\textbf{Global:}} 

\floatname{algorithm}{Procedure}

\begin{algorithm}[H]
\caption{``\texttt{SerIdx}": \ \ \ Serial index fixing of connected R-monomial with minimization}

\begin{algorithmic}[1]

\ENSURE $p, J, K, I[K], Q_F[K], Q_D[K], V[K]$.

\STATE 
Set $w:=\emptyset$.

{\bf for} $i$ from 1 to $\#Q_D[K]$, let $Q_D[K][i]$ be the $i$-th element of $Q_D[K]$, {\bf do} 

\hskip 0.6cm If $Q_D[K][i]$ has any index in $V[K]$, set $w:=w\cup \{\hbox{pre-normal form of }Q_D[K][i]\}$.

{\bf end for}

Set $Q_F[K]:=Q_F[K]$ appended by the sorted $w$ in increasing order. 

Set $Q_D[K]:=Q_D[K]\backslash w$.

\STATE 

{\bf if} $Q_F[K]$ has any index not in $V[K]$, 
let $f_i$ denote an element of $V[K]$, let $b_j$ denote an index not in $V[K]$, then

\hskip 0.6cm Set $X:=$ first element of $Q_F[K]$ having index not in $V[K]$.

\hskip 0.6cm {\bf switch} $(X)$

\hskip 0.6cm {\bf case } $\lambda R(f_1f_2,f_3b_1)$ or $\lambda R(f_1b_1,f_2f_3)${\bf :} 

\hskip 1.2cm Set $Q_F[K]:=Q_F[K]$ after renaming $b_1$ as $I[K]$.

\hskip 1.2cm Set $Q_D[K]:=Q_D[K]$ after the same renaming.

\hskip 1.2cm Set $V[K]:=V[K]\cup \{I[K]\}$, set $I[K]:=I[K]+1$.

\hskip 0.6cm  {\bf case $\lambda R(f_1b_1,f_2b_2)$:} 

\hskip 1.2cm  Set $Q_F[K]:=Q_F[K]$ after renaming $b_1, b_2$ as $I[K], I[K]+1$ respectively.

\hskip 1.2cm Set $Q_D[K]:=Q_D[K]$ after the same renaming.

\hskip 1.2cm 
Set $V[K]:=V[K]\cup \{I[K], I[K]+1\}$, set $I[K]:=I[K]+2$.

\hskip 0.6cm  {\bf case $\lambda R(f_1f_2,b_2b_3)$:} 

\hskip 1.2cm Execute procedure \texttt{Branch}.

\hskip 0.6cm  {\bf case $\lambda R(f_1b_1,b_2b_3)$:} 

\hskip 1.2cm  Set $Q_F[K]:=Q_F[K]$ after renaming $b_1$ as $I[K]$.

\hskip 1.2cm Set $Q_D[K]:=Q_D[K]$ after the same renaming.

\hskip 1.2cm 
Set $V[K]:=V[K]\cup \{I[K]\}$, set $I[K]:=I[K]+1$.

\hskip 1.2cm Execute procedure \texttt{Branch}.

\hskip 0.6cm {\bf end switch}

{\bf end if}

\STATE 

{\bf if}  $Q_F[K]\succ p$ then 

\hskip 0.6cm Set $K:=K+1$.
 
{\bf else if} either $Q_D[K]$ has any index in $V[K]$ or $Q_F[K]$ has any index not in $V[K]$,  then

\hskip 0.6cm
Execute \texttt{SerIdx}.

{\bf else}  \ \  ($Q_D[K]=\emptyset$ by the connectedness)

\hskip 0.6cm
Set $Q_F[K]:=Q_F[K]$ after the coefficients of its elements are multiplied.

\hskip 0.6cm
{\bf if} $K=1$,  then
 set $p:=Q_F[1]$. 

\hskip 0.6cm
{\bf else if} $p=-Q_F[K]$,  then
 set $p:=0$, and exit. 

\hskip 0.6cm
{\bf else}
 set $p:=\min_\prec(p, Q_F[K])$.

\hskip 0.6cm
{\bf end if}

\hskip 0.6cm
Set $K:=K+1$.

{\bf end if}

\STATE
If $K\leq J$  then
execute \texttt{SerIdx},
else exit.

\end{algorithmic}
\end{algorithm}

\floatname{algorithm}{Procedure}

\begin{algorithm}[H]
\caption{``\texttt{Branch}": \ \ \ Branched Renaming of middle R-factor}

\begin{algorithmic}[1]

\ENSURE $J, K, I[K], Q_F[K], Q_D[K], V[K]$.

\STATE 
Set $b_3, b_4:=$ 3rd, 4th index of $X$ respectively.

Set $Y_3, Y_4:=$ R-factor in $Q_F[K]\cup Q_D[K]$ other than $X$ that contains $b_3, b_4$ respectively.

\STATE
{\bf if} $Y_3\in Q_F[K]$ and $Y_3\prec Y_4$, then 

\hskip 0.6cm  Set $Q_F[K]:=Q_F[K]$ after renaming $b_3, b_4$ as $I[K], I[K]+1$ respectively.

\hskip 0.6cm Set $Q_D[K]:=Q_D[K]$ after the same renaming.

{\bf else if} $Y_4\in Q_F[K]$ and $Y_4\prec Y_3$, then 

\hskip 0.6cm
Set $Q_F[K]:=Q_F[K]$ after renaming $b_4, b_3$ as $I[K], I[K]+1$ respectively.

\hskip 0.6cm Set $Q_D[K]:=Q_D[K]$ after the same renaming.

{\bf else}\ \ \ (both $Y_3, Y_4$ are in $Q_D[K]$)

\hskip 0.6cm  Set $Q_F[K]:=Q_F[K]$ after renaming $b_3, b_4$ as $I[K], I[K]+1$ respectively.

\hskip 0.6cm Set $Q_D[K]:=Q_D[K]$ after the same renaming.

\hskip 0.6cm Set $w:=Q_F[K]$ after renaming $b_4, b_3$ as $I[K], I[K]+1$ respectively.

\hskip 0.6cm {\bf if} $w\preceq p$ then 

\hskip 1.2cm
Set $J:=J+1$.

\hskip 1.2cm
Set $Q_F[J]:=w$, set $Q_D[J]:=Q_D[K]$ after the same renaming.

\hskip 1.2cm 
Set $V[J]:=V[K]\cup \{I[K], I[K]+1\}$, set $I[J]:=I[K]+2$.

\hskip 0.6cm {\bf end if}
 
{\bf end if}

\STATE
Set $V[K]:=V[K]\cup \{I[K], I[K]+1\}$, set $I[K]:=I[K]+2$.

\end{algorithmic}
\end{algorithm}

{\it Complexity analysis:}  

\bl \label{lem:pnom}
Let $N$ be the total number of branches generated in \verb"pnom"$(f)$, where
$f$ is a connected R-monomial of degree $n$.
then $N=O(n 2^n)$.
\el

{\it Proof}. 
Let the number of free, dummy Ricci, complete, R-factors
in $f$ be $e, l, c$, respectively. Then $e+l+c=n$. The following are trivial facts:
\bi
\item If $Y$ is a free R-factor, then it generates at most two branches in both \texttt{pnom} and 
\texttt{SerIdx}; for example, this can happen
when $Y=\lambda R(f_1f_2,b_2b_3)$ or $\lambda R(f_1b_1,b_2b_3)$, 
where the $f_i$ are fixed indices and the $b_j$ are old dummy indices.

\item If $f$ has no free R-factor but has Ricci ones, then in \texttt{pnom}, the leading 
Ricci R-factor has $2l$ options, while in \texttt{SerIdx} a Ricci R-factor never generates any new branch.

\item If $f$ has only complete R-factors, then in \texttt{pnom}, the leading 
complete R-factor has $24 c$ options, while in \texttt{SerIdx} a complete R-factor
generates at most one more branch, just as a free R-factor does.
\ei

When $e\neq 0$, then
$N\leq 2^e\times 2^c=O(2^n)$.
When $e=0$ but $l\neq 0$, then $N\leq (2l)\times 2^c=O(n 2^{n})$;
when $e=l=0$, then $N\leq (24c)\times 2^{c-1}=n 2^{n-1}$.
\hfill $\square$

\vskip .2cm
In \texttt{pnom}, 
generating a complete branch takes $O(n)$ operations. By Lemma \ref{lem:pnom}, the complexity of 
\verb"pnom" is $O(n^2 2^n)$.

\section{Algorithm for Normal Form}

The {\it normal form} of an R-monomial $f$ is an R-polynomial whose 
index sequence is the R-normal form of the detailed graph of $f$. For an R-polynomial, its 
{\it normal form} is the linear combination of the normal forms of its terms.

The {\it extension} of $f$, denoted by $Ext(f)$, is an R-polynomial whose detailed graph is the
detailed extension of the detailed graph of $f$. All the R-monomials 
having the same connection multigraph with $f$ are terms of $Ext(f)$ up to coefficient.

To see how $Ext(f)$ can help computing $normal(f)$, let us check a well-known example \cite{Fulling92}.

\vskip 0.2cm
{\it Example 4.}  Let $f=R(12,34)R(13,24)$; it is already in pre-normal form.

First, $f=D$ in (\ref{pre:set}) where $n=2$, $r=0$, and $v_1=v_2=R$. So
\be\ba{ll}
& \pnom(Ext(f)) \\

=& \pnom\Big(\lambda_0 (\lambda_{12} R(12,34) + \lambda_{13} R(13,42) +\lambda_{14} R(14,23))\Bigstrut\\

& \phantom{\pnom\Big(\lambda_0}
(\lambda_{22} R(13,24) + \lambda_{23} R(12,43) +\lambda_{24} R(14,32))\Big) \\

=& \lambda_0 (\lambda_{12} \lambda_{22}
+\lambda_{12} \lambda_{24}
+\lambda_{13} \lambda_{23}
+\lambda_{13} \lambda_{24}
+\lambda_{14} \lambda_{22}
+\lambda_{14} \lambda_{23}
) R(12,34)R(13,24) \Bigstrut\\

& \hfill -\lambda_0(\lambda_{12} \lambda_{23}
+\lambda_{13} \lambda_{22}
+\lambda_{14} \lambda_{24}
)
R(12,34)R(12,34).
\ea
\label{ex4:ext}
\ee

Let $x_1=R(12,34)R(13,24)=f$ and $x_2=R(12,34)R(12,34)$. Then $x_1\succ x_2$ in lexicographic order.
Setting $\lambda_0=\lambda_{12}=\lambda_{13}=\lambda_{14}=1$ in (\ref{ex4:ext}), we get the following Bianchi relation
on the first R-factor of $f$:
\[
2(\lambda_{22}+\lambda_{23}+\lambda_{24})x_2-(\lambda_{22}+\lambda_{23}+\lambda_{24})x_1=0.
\]
Setting $\lambda_0=\lambda_{22}=\lambda_{23}=\lambda_{24}=1$ in (\ref{ex4:ext}), we get
the following Bianchi relation
on the second R-factor of $f$:
\[
2(\lambda_{12}+\lambda_{13}+\lambda_{14})x_2-(\lambda_{12}+\lambda_{13}+\lambda_{14})x_1=0.
\]
Solving the two equations in variables $x_1, x_2$, we get the solution $x_2=x_1/2$.
So $x_1/2$ is the normal form of $f$.
\vskip 0.2cm

Example 4 suggests the following procedure of normal form computing:
Let connected R-monomial $f$ take the form (\ref{ex:loop:no}) where every $v_i$ represents $R$, then
the pre-normal form of its extension (\ref{pre:set}) is evaluated to zero $n$ times
under $n$ special evaluations of the parameters $\lambda$'s: for $i=1..n$, the $i$-th evaluation
is by setting $\lambda_r=\lambda_{i2}=\lambda_{i3}=\lambda_{i4}=1$. Denote by $\Lambda$ 
the set of parameters $\lambda$'s in $Ext(f)$ other than $\lambda_r$. Then 
$\# \Lambda=3^{n-r}$.
Let there be $m$ different pre-normal R-monomials in $\pnom(Ext(f))$. 
Denote them by $x_1\succ x_2\succ  \cdots \succ  x_m$ 
following the lexicographic order. That $\pnom(Ext(f))=0$ under
the above $n$ special evaluations gives $n$ linear equations 
\[
\A\x=0, 
\]
where 
$\x=(x_1, x_2, \ldots, x_m)^T$, and $\A_{n\times m}$ is a matrix whose entries are in ${\mathbb Q}[\Lambda]$.
Solving the linear system by Gauss-Jordan elimination, one gets the {\bf RREF} $\Rref_1$:
$x_j+\sum_{k>j} \mu_k x_k=0$ for $j\in I\subseteq \{1,2,\ldots, m\}$, 
where $\mu_k\in {\mathbb Q}(\Lambda)$. 

In each equation of $\Rref_1$, divide the expression on the left side into two 
sub-expressions: sub-expression 1 contains the terms with coefficient in $\mathbb Q$, 
and sub-expression 2 contains the rest. 
Denote by $sub_{\mathbb Q}(\Rref_1)$ the union of the sub-expression 1's from the equations of $\Rref_1$,
and denote by $sub_{\check{\mathbb Q}}(\Rref_1)$ the union of the sub-expression 2's.
Then each element of $sub_{\check{\mathbb Q}}(\Rref_1)$ must be equal to zero, and we get
a set of at most $\# I$ linear equations in the variables $x_l$ where 
$l$ is in a subset of $\{1,2,\ldots, m\}\backslash I$.
Computing the RREF of this linear system, we get $\Rref_2$. 
Continuing the selection of terms with coefficient $\notin \mathbb Q$ in $\Rref_i$ and the computing
of an RREF $\Rref_{i+1}$ of the selected new linear system, we finally get a complete
RREF with coefficient in $\mathbb Q$, which is composed of the
$sub_{\mathbb Q}$'s in the $\Rref$'s.

\bp
The complete RREF obtained by the above procedure is the RREF of 
linear system (\ref{triangulate}).
\ep

{\it Proof.} 
Let $\C\x=0$ be the complete RREF obtained from $\Rref_1, \Rref_2, \ldots, \Rref_N$.
Then any equation in $\Rref_N$ has all the coefficients in $\mathbb Q$, {\it i.e.}, 
$sub_{\check{\mathbb Q}}(\Rref_N)=\emptyset$.
Since $\Rref_N$ is obtained from $sub_{\check{\mathbb Q}}(\Rref_{N-1})=0$ by elementary row transformations with coefficients
in ${\mathbb Q}(\Lambda)$, 
every equation of $sub_{\check{\mathbb Q}}(\Rref_{N-1})$ is a ${\mathbb Q}(\Lambda)$-linear combination of  
the equations in $\Rref_N$. 
We use $sub_{\check{\mathbb Q}}(\Rref_{N-1})\subseteq 
\langle sub_{\mathbb Q}(\Rref_{N}) \rangle_{{\mathbb Q}(\Lambda)}$ to denote this relation.

Similarly,
$sub_{\mathbb Q}(\Rref_{N-1})$ also has all its coefficients in $\mathbb Q$, and
\[
sub_{\check{\mathbb Q}}(\Rref_{N-2})\subseteq 
\Big\langle sub_{\mathbb Q}(\Rref_{N-1}), sub_{\check{\mathbb Q}}(\Rref_{N-1}) \Big\rangle_{{\mathbb Q}(\Lambda)}
\subseteq 
\Big\langle sub_{\mathbb Q}(\Rref_{N-1}), sub_{\mathbb Q}(\Rref_{N}) \Big\rangle_{{\mathbb Q}(\Lambda)}
\]
Continuing this argument, we get
\be
\hbox{all rows of }\A 
\subseteq 
\Big\langle sub_{\mathbb Q}(\Rref_{1}), sub_{\check{\mathbb Q}}(\Rref_{1}) \Big\rangle_{{\mathbb Q}(\Lambda)}
\subseteq  \ldots 
\subseteq  \ds
\Big\langle \bigcup_{i=1}^N sub_{\mathbb Q}(\Rref_{i}) \Big\rangle_{{\mathbb Q}(\Lambda)}.
\label{proof:end}
\ee

Let $\B\x=0$ be a linear system obtained from $\A\x=0$ by $M\geq 3^{n-r}$ different
generic ${\mathbb Q}$-specifications of the parameters in $\Lambda$. 
Then $\B$ has $nM$ rows. Let $\D\x=0$ be the linear system (\ref{triangulate}) of 
$3^n-2^n$ rows. 
Denote by $\langle \P \rangle_{\mathbb Q}$ the row space of a matrix $\P$. 

Then
$\langle \B \rangle_{\mathbb Q}=\langle \D \rangle_{\mathbb Q}$. 
By (\ref{proof:end}), $\langle \B \rangle_{\mathbb Q}\subseteq 
\langle \C \rangle_{\mathbb Q}$. So $\langle \D \rangle_{\mathbb Q}\subseteq 
\langle \C \rangle_{\mathbb Q}$.

Conversely, let $\E$ be the RREF of $\D$, let the equations in $\E\x=0$ be
$x_j+\sum_{k>j} \mu_k x_k=0$ for $k\in J\subseteq \{1,2,\ldots, m\}$, 
where $\mu_k\in \mathbb Q$. 
Then for all $j\notin J$, $x_j\in \QJ$ and the linear dependency 
is represented explicitly by a row of $\E$.

Let there be an equation $Eq\in \Rref_1$ in which 
$sub_{\check{\mathbb Q}}\neq 0$. Then $Eq$ is of the form
$
c+f+g/h=0,
$
where \\
(1) $c=sub_{\mathbb Q}\in \QJ$ and is linear in the $x_l$ for $l\in J$;\\
(2) $f,g\in \QJ[\Lambda]$ and are both linear in the $x_l$ for $l\in J$; \\
(3) $h\in {\mathbb Q}[\Lambda]$; \\
(4) either $f=0$ or $f$ has no term in $\QJ$;\\
(5) either $g=0$ or no term of $g$ can be divided by $h$; \\
(6) $sub_{\check{\mathbb Q}}=f+g/h\neq 0$.

Using the rows of $\E$ to make elementary row transformations to $Eq$, it is easy to see that
$Eq$ is changed into
\[
\sum_{k\in J} (\gamma_k+\alpha_k+\frac{\beta_k}{h})x_k=0,
\]
where \\
(1) $c=\sum_{k\in J} \gamma_k x_k$, \
$f=\sum_{k\in J} \alpha_k x_k$, \
$g=\sum_{k\in J} \beta_k x_k$;\\
(2) $\gamma_k\in \mathbb Q$;\\
(3) either $\alpha_k=0$ or 
every term of $\alpha_k\in {\mathbb Q}[\Lambda]$ has degree $>0$;\\
(4) either $\beta_k=0$ or no term of $\beta_k\in {\mathbb Q}[\Lambda]$ can be divided by $h$.

We prove that for all $k\in J$, $\alpha_k=\beta_k=0$. 

When $Eq\in \Rref_1$ is replaced by the corresponding $M$ equations $Eq_i\in 
\Rref_1(\A|_i)$ for $i=1..M$, we get $M$ equations 
\be
\sum_{k\in J} (\gamma_k+\alpha_k|_i+\frac{\beta_k|_i}{h|_i})x_k=0,  \label{proof:eqi}
\ee
where $h|_i=h$ by 
generic specification $i$, and so for the $\alpha_k|_i, \beta_k|_i$. These equations can all be obtained from $\E\x=0$ by
$\mathbb Q$-coefficient elementary row transformations.

If for some $k\in J$, $\alpha_k+\beta_k/h\neq 0$, then
in the $(3^{n-r}+1)$-dimensional vector space with coordinates $(\Lambda, y)$, 
hyperplane $y=-\gamma_k$ generically does not meet hypersurface
$y=\alpha_k+\beta_k/h$. This means 
we can choose a generic specification $i$ such that 
$\gamma_k+\alpha_k|_i+(\beta_k|_i)/(h|_i)\neq 0$.
Under this generic specification, (\ref{proof:eqi}) 
 is a nontrivial linear relation among the $x_l$ where $l\in J$. 
This violates the RREF property of $\D$ that 
the $\{x_k\,|\,k\in J\}$ are $\mathbb Q$-linearly independent.

So $\alpha_k+\beta_k/h$ is identical to zero for all $k\in J$, hence $f+g/h$
is identical to zero. This means $c=0$ is obtained from $Eq$
by applying some ${\mathbb Q}(\Lambda)$-coefficient elementary row transformations induced by $\E$.

Applying this argument to all equations of 
$\Rref_1$, we get that all elements of $sub_{\mathbb Q}(\Rref_1)$ and 
$sub_{\check{\mathbb Q}}(\Rref_1)$ can be obtained from $\D$
by ${\mathbb Q}(\Lambda)$-coefficient elementary row transformations.

Continue this argument to $\Rref_2$, then inductively to
all $\Rref_j$ for $j>1$, till
$j=N$. In the end, $\cup_{j=1}^N sub_{\mathbb Q}(\Rref_j)$ can be obtained from $\D$ by
${\mathbb Q}(\Lambda)$-coefficient elementary row transformations.

Now make $M$ generic specifications of $\Lambda$ to turn the 
${\mathbb Q}(\Lambda)$-coefficient elementary row transformations into 
${\mathbb Q}$-coefficient ones. We finally get $\langle \C \rangle_{\mathbb Q}\subseteq 
\langle \D \rangle_{\mathbb Q}$.
\hfill $\square$
\vskip .2cm

As a corollary, 
if $\pnom(f)$ is up to coefficient a leading variable $x_j$ in an equation
$x_j+\sum_{k>j} \lambda\mu_k x_k=0$ of the complete RREF, say
$\pnom(f)=\lambda x_j$, then the normal form of $f$ is
$\sum_{k>j} (-\lambda\mu_k) x_k$, else $\pnom(f)$ is the normal form of $f$.

\renewcommand{\algorithmicensure}{\textbf{Output:}} 

\floatname{algorithm}{Function}

\begin{algorithm}[H]
\caption{``\texttt{Rebe}":\ \ \
{\bf R}educed row {\bf e}chelon form of {\bf B}ianchi relations in the {\bf e}xtension of a connected R-monomial}

\begin{algorithmic}[1]
\REQUIRE
$f$, a connected R-monomial without integer indices.

\ENSURE
a $\mathbb Q$-linear RREF in pre-normal R-monomials. 

\STATE 
Set $g:=\texttt{pnom}(Ext(f))$ after like term combination.

Set $h:=$ sorted ascending sequence of the monic R-monomials of $g$.

Set $m:=\#$ elements of $h$. 


\STATE 
Set $H:=\emptyset$.

{\bf for} $j=1$ to $n$ {\bf do}

\hskip .6cm Set $w:=g$ after the evaluation
$\lambda_r=\lambda_{j2}=\lambda_{j3}=\lambda_{j4}=1$.

\hskip .6cm Set $H:=H\cup \{w=0\}$.

{\bf end for}

Set $H:=$ coefficient matrix of $H$ in variables $h[1]\succ \ldots\succ h[m]$,
where $h[i]$ denotes the $i$-th element of $h$.

\STATE 

Set $\C:=\emptyset$, set $q:=0$.

{\bf do}

\hskip .6cm Set $Q:=$ RREF of $H$, where rows having zero only are removed.

\hskip .6cm Set $a, b:=$ number of rows, columns of $Q$ respectively.

\hskip .6cm Set $k:=$ column number of the first column of $Q$ having entry $\notin \mathbb Q$.

\hskip .6cm {\bf if} $k\leq b$ then

\hskip 1.2cm Set $H:=$ matrix of zeroes with size $a\times (b-k+1)$.

\hskip 1.2cm {\bf for} $i=1$ to $a$, {\bf do}

\hskip 1.8cm {\bf for} $j=k$ to $b$,
let $Q[i,j]$ be the $(i,j)$ entry of $Q$, {\bf do}

\hskip 2.4cm Set $w:=$ the term of $Q[i,j]$ in $\mathbb Q$. 

\hskip 2.4cm Set $Q[i,j]:=w$, set
$H[i,j-k+1]:=Q[i,j]-w$.

\hskip 1.8cm {\bf end for} 

\hskip 1.2cm {\bf end for}

\hskip .6cm {\bf end if}

\hskip .6cm Set $Q:=Q$ augmented on the left by the $a\times q$ matrix of zeroes. 

\hskip .6cm Set $\C:=\C$ after appending $Q$.

\hskip .6cm Set $q:=q+k-1$.

{\bf while} $k\leq b$.

\RETURN  $\C(h[1], \ldots, h[m])^T=0$.

\end{algorithmic}
\end{algorithm}

\floatname{algorithm}{Algorithm}

\begin{algorithm}[H]
\caption{``\texttt{normal}":\ \ \
Normal form of R-polynomial.
}

\begin{algorithmic}[1]
\REQUIRE 
$f$, an R-polynomial without integer indices.

\ENSURE
The normal form of $f$.

\STATE 

Set $U, L:=$ set of upper, lower free indices in $f$ respectively.

\STATE

(pre-normalizing maximal connected R-submonomials)

Set $C:=$ maximal connected monic R-submonomials in the terms of $f$.

{\bf for} $i=1$ to $\# C$, let $C[i]$ be the $i$-th element of $C$, {\bf do}

\hskip 0.6cm 
Set $D[i]:=\texttt{pnom}(C[i])$.

\hskip 0.6cm 
Set $f:=f$ after replacing $C[i]$ with $D[i]$ and combining like terms.

{\bf end for}

If $f=0$ then return $0$ and exit.

\STATE 

(normalizing maximal connected R-submonomials)

Set $C:=$ maximal monic connected R-submonomials in the terms of $f$. 

Set $D:=\emptyset$.

{\bf do}
 
\hskip 0.6cm {\bf if} ${\rm deg}(C[1])=1$, then

\hskip 1.2cm Set $D:=D\cup \{C[1]=C[1]\}$, set $C:=C\backslash \{C[1]\}$.

\hskip 0.6cm {\bf else}

\hskip 1.2cm 
Set $H:=\texttt{Rebe}(C[1])$.

\hskip 1.2cm 
{\bf for} $i=1$ to $\# C$, {\bf do}

\hskip 1.8cm {\bf if} $C[i]$ is a leading variable of $H$, then 

\hskip 2.4cm Set $w:=C[i]$ after applying $H$ as elimination rules.

\hskip 2.4cm Set $D:=D\cup \{C[i]=w\}$,  set $C:=C\backslash \{C[i]\}$.

\hskip 1.8cm {\bf else if} $C[i]$ is a variable of $H$, then 

\hskip 2.4cm Set $D:=D\cup \{C[i]=C[i]\}$, set $C:=C\backslash \{C[i]\}$.

\hskip 1.8cm {\bf end if}

\hskip 1.2cm {\bf end for}

\hskip 0.6cm {\bf end if}

{\bf while} $C\neq \emptyset$.

\STATE 

(ordinary polynomial operations; all terms are sorted, so are the 
maximal connected R-submonomials within each term)

Set $f:=f$ after applying $D$ as substitution rules, making linear expansion and
combining like terms.

If $f=0$ then return $0$ and exit.

\STATE 

(renaming integer indices in each term)

{\bf for} $i=1$ to $\#$ terms of $f$, let $f[i]$ be the $i$-th term, {\bf do}

\hskip 0.6cm 
Set $C:=$ sequence of maximal connected R-submonomials of $f[i]$.

\hskip 0.6cm {\bf if} $\#C\neq 1$ then 

\hskip 1.2cm Set $w[0]:=0$.

\hskip 1.2cm {\bf for} $j=1$ to $\#C$ {\bf do}

\hskip 1.8cm Set $w[j] :=$ maximal integer index in $C[j]$.

\hskip 1.8cm For $k=1$ to $w[j]$, 
replace dummy index $k$ in $C[j]$ with dummy index $w[j-1]+k$.

\hskip 1.8cm Set $w[j]:=w[j]+w[j-1]$.

\hskip 1.2cm {\bf end for}

\hskip 0.6cm {\bf end if}

\hskip 0.6cm Set $f[i]:=C$.

{\bf end for} 

\STATE

Restore the row characters of the indices in $f$: free indices are restored according to their
records $U$ and $L$; a dummy index in its first/second occurrence of a term is set to the 
upper/lower row respectively.














\RETURN $f$.

\end{algorithmic}
\end{algorithm}


{\it Complexity analysis}:

\bp
Let $f$ be a connected R-monomial of degree $n$. 
Then \verb"normal"$(f)$ takes $O(n 9^n)$ operations.
\ep

{\it Proof.} 
Let $f$ be in the form  of (\ref{ex:loop:no}), then
$Ext(f)$ has $3^{n-r}$ terms. Computing the pre-normal form of $Ext(f)$ takes $O(n^2 6^n)$ operations. 

Let $m$ be the number of R-monomials in $\pnom(Ext(f))$ after like term combination. Then $m=O(3^n)$.
In \verb"Rebe", computing the RREF $\Rref_1$ with coefficient field ${\mathbb Q}(\Lambda)$ 
takes $O(n^2m)$ arithmetic operations upon multivariate rational functions. 

Let $r={\rm rank}(\Rref_1)$,
then $sub_{\check{\mathbb Q}}(\Rref_1)$ when written as a matrix has the size of $r\times (m-r)$ at most, so computing 
$\Rref_2$ takes $O(r^2(m-r))$ operations, which is $O(n^2(m-n))$ when $m\gg n$.
Going this way, computing $\Rref_{i+1}$ takes $O(n^2(m-in))$ operations, till $i=[m/n]$.
Since
\be
\sum_{i=0}^{[m/n]} n^2 (m-in)
=n^2m([m/n]+1)-n^3[m/n]([m/n]+1)/2
=O(mn(m+n)),
\ee
computing the complete RREF takes $O(nm^2)$ operations, which is $O(n 9^n)$ when $m=\Theta(3^n)$. 
The overall complexity is thus $O(n 9^n)$.
\hfill $\square$

\vskip .2cm
Function \verb"Rebe" in \verb"pnom" can be replaced by other methods for computing 
the RREF of (\ref{triangulate}).
Direct solving of (\ref{triangulate}) by Gauss-Jordan elimination has complexity
$O(m9^n)$. Although it is not fair to take the complexity of an arithmetic operation 
on rational functions as the same with that on rational numbers,
function \verb"Rebe" reduces the equation-solving complexity to $O(nm^2)$, which is polynomial in the case when
$m$ is the size of a polynomial in $n$. 

Notice that every row of (\ref{triangulate}) has at most three nonzero entries.
It may be possible that the iterative methods for sparse $\mathbb R$-linear systems be applied to 
the sparse $\mathbb Q$-linear system (\ref{triangulate}) for
infinitely many {\bf accurate} $\mathbb Q$-valued solutions by solving the corresponding  normal equations \cite{Saad03}, so that 
the sparse solving has complexity $O(3^n)$. 
Even so, \verb"Rebe" is valuable in the case when $m=O(3^{n/2})$.

\section{Conclusion}

In this paper we establish the graph algebra extension theory and develop an algorithm of 
normalizing Riemann tensor polynomials based on this theory. The theory can be
extended to the case involving covariant derivatives of the Riemann tensor, and other 
types of tensor in a straightforward way.
Future work includes such extensions, and application to theorem 
proving in Riemannian geometry.

\end{document}